# Laser-induced terahertz spin transport in magnetic nanostructures arises from the same force as ultrafast demagnetization

S.M. Rouzegar[1,2], L. Brandt[3], L. Nádvorník[1,2,4], D.A. Reiss[1], A.L. Chekhov[1,2], O. Gueckstock[1,2], C. In[1,2], M. Wolf[2], T.S. Seifert[1], P.W. Brouwer[1], G. Woltersdorf[3], T. Kampfrath[1,2]

1. Department of Physics, Freie Universität Berlin, 14195 Berlin, Germany
2. Department of Physical Chemistry, Fritz Haber Institute of the Max Planck Society, 14195 Berlin, Germany
3. Institut für Physik, Martin-Luther-Universität Halle, 06120 Halle, Germany
4. Faculty of Mathematics and Physics, Charles University, Ke Karlovu 3, 121 16 Prague, Czech Republic

**Abstract**

Laser-induced terahertz spin transport (TST) and ultrafast demagnetization (UDM) are central but so far disconnected phenomena in femtomagnetism and terahertz spintronics. Here, we use broadband terahertz emission spectroscopy to reliably measure both processes in one setup. We find that the rate of UDM of a single ferromagnetic metal film F has the same time evolution as the flux of TST from F into an adjacent normal-metal layer N. This remarkable agreement shows that UDM and TST are driven by the same force, which is fully determined by the state of the ferromagnet. An analytical model consistently and quantitatively explains our observations. It reveals that both UDM in F and TST in the F|N stack arise from a generalized spin voltage, which is defined for arbitrary, nonthermal electron distributions. We also conclude that contributions due to a possible temperature difference between F and N are minor and that the spin-current amplitude can, in principle, be increased by one order of magnitude. In general, our findings allow one to apply the vast knowledge of UDM to TST, thereby opening up new pathways toward large-amplitude terahertz spin currents and, thus, energy-efficient ultrafast spintronic devices.

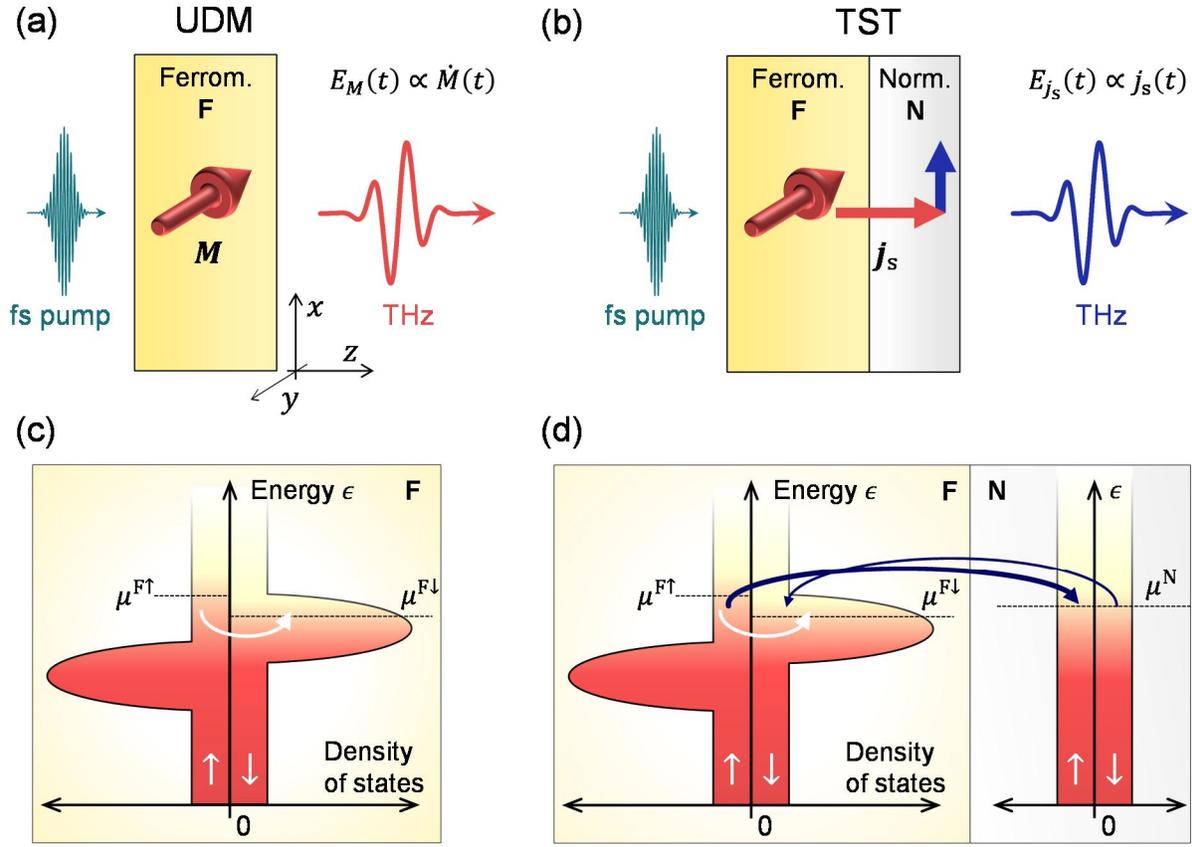

FIG 1. Ultrafast demagnetization (UDM) vs terahertz spin transport (TST). (a) Side view of a single ferromagnetic metal layer (F) with magnetization $\boldsymbol{M} = M\boldsymbol{u}_y$ parallel to the $y$ axis with unit vector $\boldsymbol{u}_y$. Excitation by a femtosecond laser pulse triggers UDM. The transient magnetic dipole gives rise to the emission of a terahertz pulse with field $E_M(t) \propto \dot{M}(t)$. (b) F|N stack consisting of F and an adjacent normal paramagnetic metal layer (N). Femtosecond laser excitation drives a spin current with density $\boldsymbol{j}_s = j_s \boldsymbol{u}_z$ from F to N. In N, $\boldsymbol{j}_s$ is converted into a charge current with density $\boldsymbol{j}_c$, leading to the emission of a terahertz electromagnetic pulse with electric field $E_{j_s}(t) \propto j_s(t)$ directly behind the sample. Both $E_M(t)$ and $E_{j_s}(t)$ are linearly polarized perpendicular to $\boldsymbol{M}$ and measured by electrooptic sampling. (Supplemental Materialc) Schematic of the density of states of spin-up (↑) and spin-down (↓) electrons of a Stoner-type ferromagnet such as Fe. Quasi-elastic spin-flip scattering events (white curved arrow) lead to transfer of spin angular momentum to the crystal lattice. (d) N acts as an additional sink of spin angular momentum through spin-conserving electron transfer across the F|N interface (blue curved arrows). In (a) and (b), the spin transfer rate scales with the generalized spin voltage $\Delta\tilde{\mu}_s$ [Eq. (6)], which equals $\mu^{F\uparrow} - \mu^{F\downarrow}$ in the case of Fermi-Dirac electron distributions.

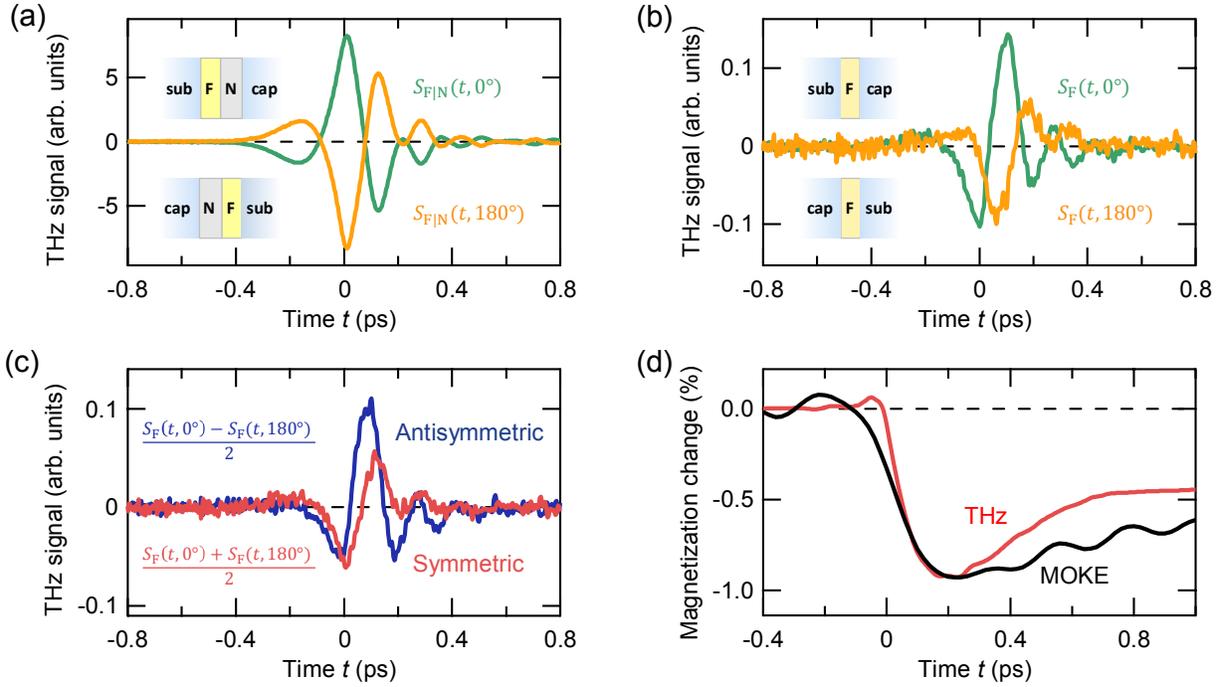

FIG. 2. Typical terahertz electrooptic signals, odd with respect to magnetization $\mathbf{M}$, from F and F|N samples consisting of F=CoFe(3 nm) and N=Pt(3 nm). (a) Terahertz emission signal $S_{F|N}(t, 0°)$ from an F|N stack. When the sample is turned by 180° about $\mathbf{M}$, the signal $S_{F|N}(t, 180°)$ is obtained. Note that the sample is optically symmetrized by a cap window that is identical to the diamond substrate. (b) Same as panel (a), but for the F sample. Note the asymmetry between $S_F(t, 0°)$ and $S_F(t, 180°)$. (c) Signals $S_F^+(t)$ and $S_F^-(t)$ symmetric and antisymmetric with respect to sample turning. (d) Extracted magnetization dynamics from the symmetric signal of panel (c) (red curve), along with magnetization dynamics as measured by the magnetooptic Kerr effect (MOKE, black curve).

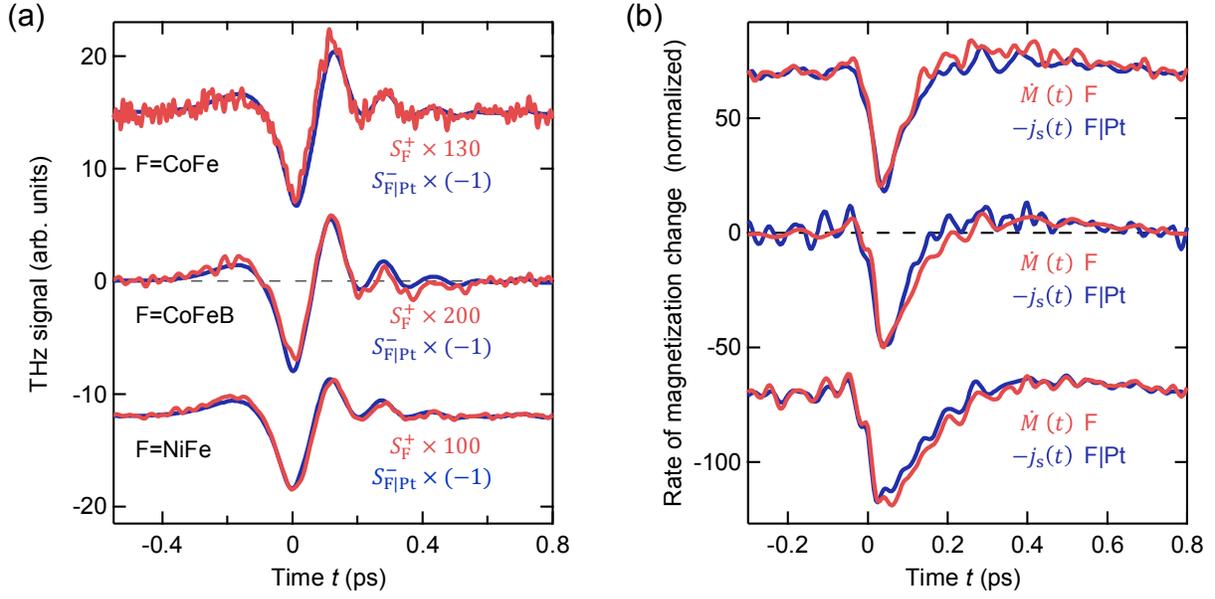

FIG. 3. Terahertz emission due to TST in F|N vs UDM in F. (a) Terahertz signal $S^-_{F|N}(t)$ from a CoFe(3 nm)|Pt(3 nm) sample, antisymmetric with respect to sample turning about $\boldsymbol{M}$ (blue solid line), vs terahertz signal $S^+_F(t)$ from a single CoFe(3 nm) layer, symmetric with respect to sample turning (red solid line). The curves below show analogous signals for CoFeB(5 nm)|Pt(3 nm), CoFeB(5 nm) and NiFe(9 nm)|Pt(3 nm), NiFe(9 nm) samples. The curves are scaled by the indicated factors and offset vertically for clarity. (b) Temporal evolution of the spin current $j_s$ flowing in the F|N sample and of the rate of change $\dot{M}$ of the F sample's magnetization as extracted from the data of panel (a). Curves are normalized to their minima.

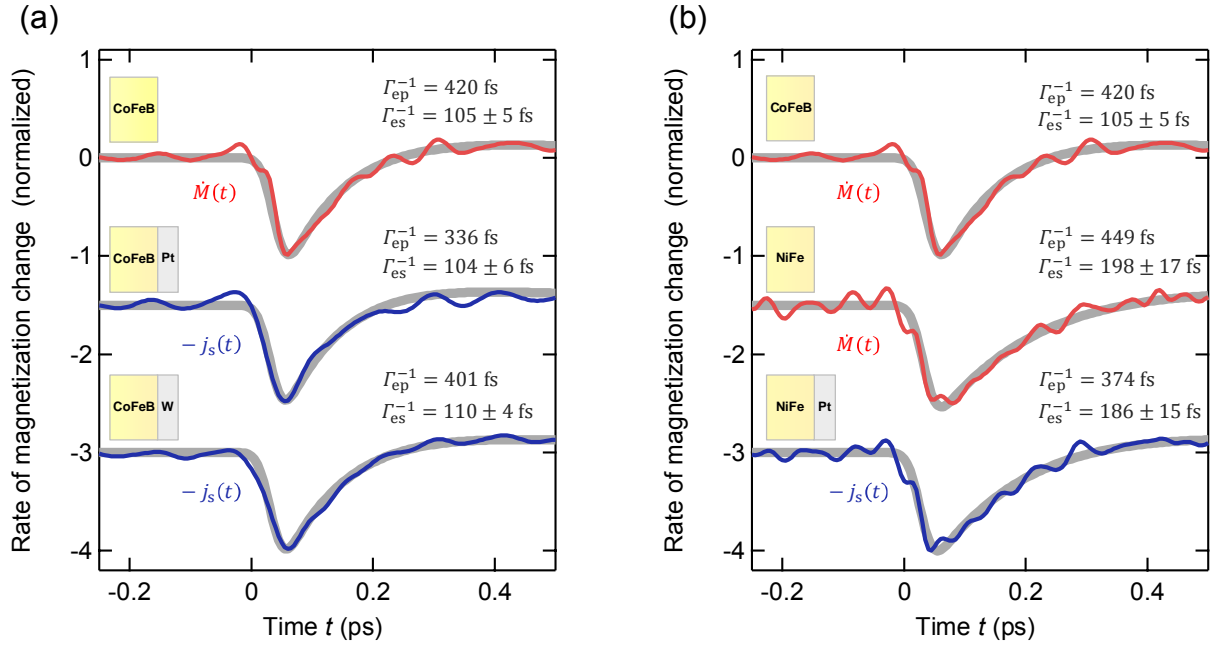

FIG. 4. Measured and modeled dynamics of $\dot{M}$ and $j_s$. (a) Measured dynamics of the rate of change $\dot{M}(t)$ of the magnetization of a CoFeB film (red solid line) and the spin current $j_s(t)$ of CoFeB|Pt and CoFeB|W stacks (blue solid lines). Grey solid lines are fits based on Eq. (10) with $\Gamma_{es}$ and the overall amplitude scaling as the only fit parameters. (b) Analogous to (a), but for CoFeB, NiFe and NiFe|Pt.

## I. INTRODUCTION

Fundamental operations in future spin-based electronics are the manipulation of magnetic order, the transport of spin angular momentum and the detection of spin dynamics.[1] The research fields of femtomagnetism and terahertz spintronics aim to push the three operations to femtosecond time scales and, thus, terahertz bandwidth.[2,3,4,5,6,7] Figure 1(a,b) shows the model systems in which two key phenomena of ultrafast spin dynamics are studied extensively.

In a single ferromagnetic metal layer F, uniform excitation by a femtosecond laser pulse induces ultrafast demagnetization [UDM; Fig. 1(a)].[2,8,9,10,11] This effect reveals the time scales of elementary spin interactions with electron orbital and lattice degrees of freedom and is a central ingredient for ultrafast magnetization switching.[3,6] Theories of UDM involve spin flips[10,12,13,14] or magnon emission[15] due to electron scattering together with spin-orbit coupling.[16]

In F|N stacks, where N is a normal paramagnetic metal layer, uniform laser excitation triggers terahertz spin transport (TST) between F and N [Fig. 1(b)].[5,6,17,18,19,20,21] Such spin currents can exert spin torque at ultrashort time and length scales. They may, thus, excite terahertz magnons[22,23] and, ultimately, switch magnetic order.[3] TST also serves to efficiently generate broadband terahertz electromagnetic pulses for photonic and spectroscopy applications.[24,25,26,27,28] Simulations show that TST is superdiffusive.[29,30,31]

Spin transport can, in general, be driven by gradients of temperature and spin voltage.[32,33,34] On one hand, a transient spin voltage was recently observed in laser-excited single layers of ferromagnetic Fe.[35] On the other hand, theoretical arguments[36,37] indicate that the spin voltage, plus temperature differences between spin-up and spin-down electrons, could drive demagnetization. It follows that the seemingly disconnected phenomena of TST and UDM may share a common driving force. Experimental evidence for this exciting conjecture is, however, missing. It is also far from obvious whether concepts like spin voltage and temperature can be applied to nonthermal electron states that prevail in the first 100 fs after optical excitation and ultimately determine the bandwidth of terahertz spintronic devices.

Here, we use terahertz emission spectroscopy to demonstrate that the rate of UDM in F samples [Fig. 1(a)] and the flux of TST in F|N stacks [Fig. 1(b)] have identical time evolution down to the 40 fs resolution of our experiment. Our measurements along with an analytical model show that UDM and TST are driven by a common dominant force: a generalized spin voltage of the electrons in F, which is defined for arbitrary, nonthermal electron distributions. These insights open up entirely new perspectives and synergies because they allow us to better understand and ultimately optimize TST by exploiting the extensive knowledge about UDM. For example, our results indicate that the temporal onset of TST is only determined by the duration of the femtosecond pump pulse and that the amplitude of TST can, in principle, be increased by one order of magnitude.

## II. EXPERIMENTAL SETUP

**Samples and excitation**

As F materials, we choose the Stoner-type ferromagnets $Co_{70}Fe_{30}$ (CoFe), $Co_{40}Fe_{40}B_{20}$ (CoFeB) and $Ni_{80}Fe_{20}$ (NiFe). For N, we choose the spin-to-charge conversion materials Pt and W because they exhibit large yet opposite spin Hall angles.[24] Two thin films of F and F|N are grown by magnetron sputtering on the same diamond substrate, which is transparent at all relevant terahertz and optical frequencies. The sample preparation is detailed in Appendix A.

The direction of the sample magnetization $M$ is set by an external magnetic field of 10 mT either parallel or antiparallel to the $y$-axis unit vector $u_y$ [Fig. 1(a)]. The samples are excited with linearly polarized laser pulses (wavelength of 800 nm, duration of 10 fs and pulse energy of 2 nJ) from a Ti:sapphire laser oscillator (repetition rate at 80 MHz) under normal incidence. The pump beam diameter at the sample position was approximately 25 μm full width at half maximum of the intensity.

## Measurement of UDM and TST

***Terahertz field emission.*** To measure the dynamics of the magnetization $\boldsymbol{M}(t) = M(t)\boldsymbol{u}_y$ of an F sample [Fig. 1(a)] and of the spin current flowing from F into an adjacent N [Fig. 1(b)] vs time $t$, the concomitantly emitted terahertz electromagnetic pulse is an excellent probe. UDM [Fig. 1(a)] implies a dynamic magnetic dipole that generates an electromagnetic pulse[38,39] with an electric field

$$E_M(t) \propto \dot{M}(t) \tag{1}$$

directly behind the sample (see Appendix A).

In TST [Fig. 1(b)], the spin-current density $\boldsymbol{j}_s(t) = j_s(t)\boldsymbol{u}_z$ across the F-N interface is instantaneously converted[40] into a transverse charge-current density proportional to $j_s(t)$ by the inverse spin Hall effect in N. It results in a time-dependent electric dipole and, thus, emission of an electromagnetic pulse with transient electric field[24,25,26,27]

$$E_{j_s}(t) \propto j_s(t) \tag{2}$$

behind the sample (see Appendix A). As the dynamics are driven by a femtosecond laser pulse, the bandwidth of $E_M$ and $E_{j_s}$ is expected to extend to frequencies well above 10 THz.

In our setup, we detect any transient electric field $E(t)$ such as $E_{j_s}$ and $E_M$ by electrooptic sampling,[41] where a probe pulse (0.6 nJ, 10 fs) copropagates with the terahertz pulse through an electrooptic crystal. The ellipticity $S(t)$ accumulated by the sampling pulse is measured as a function of the delay $t$ between terahertz and sampling pulse by means of a polarization-sensitive optical bridge, which consists of a quarter-wave plate, a polarizing beam splitter and two balanced photodiodes. As electrooptic crystal, we use GaP(110) (thickness of 250 µm) if not explicitly mentioned otherwise, but also ZnTe(110) (thickness of 1 mm). All experiments are performed at room temperature in a dry $N_2$ atmosphere.

***From signals to fields.*** To focus on magnetic effects, we only consider the signal component $S(t) = [S(t, \boldsymbol{M}) - S(t, -\boldsymbol{M})]/2$. The waveform $S(t)$ is connected to the terahertz electric field $E(t)$ directly behind the sample by the convolution[42]

$$S(t) = (H_{SE} * E)(t) = \int d\tau\, H_{SE}(t - \tau) E(\tau). \tag{3}$$

The transfer function $H_{SE}(t)$ mediates between $S$ and $E$ and accounts for the THz pulse propagation to the detection and the electrooptic-sampling process.[41] We determine $H_{SE}$ by using an appropriate reference emitter.[43] By numerical inversion of Eq. (3), $E(t)$ and, thus, $\dot{M}(t)$ [Eq. (1)] and $j_s(t)$ [Eq. (2)] are obtained with an estimated time resolution of 40 fs.

***Expected signal contributions.*** The total THz field behind the F|Pt and F|W stacks is dominated[24] by $E_{j_s}$. Due to its electric-dipole character, $E_{j_s}(t)$ fully reverses when the F|N stack is turned by 180° about an axis parallel to $\boldsymbol{M}$ [Fig. 1(b)]. In contrast, the field $E_M$ from the F sample originates from magnetic dipoles. It, thus, remains invariant under 180° sample turning [Fig. 1(a)] and is much smaller[39] than $E_{j_s}$.

To minimize competing signals due to pump-induced gradients,[44] the F thickness is chosen sufficiently small. To discriminate electric-dipole signals due to a possible inversion asymmetry of the F sample,[45] we measure it both in the 0° and the 180°-turned configuration. For this purpose, the samples are macroscopically symmetrized by adding a cap layer (cap) that is identical to the substrate [sub; see inset of Fig. 2(a)]. Details of this separation procedure and two complementary approaches are described in Appendix A and Note 1 of the Supplemental Material.

***Magneto-optic probing.*** For comparison to UDM probed by terahertz spectroscopy [Eq. (1)], we also conduct a pump-probe experiment in which $M(t)$ of the F sample is measured by the transient magnetooptic Kerr effect (MOKE, see Appendix A).

## III. RESULTS

**Terahertz emission signals**

**F|N sample.** Figure 2(a) shows the signal $S(t)$ from a sub||F|N||cap sample where F=CoFe(3 nm) and N=Pt(3 nm). As expected from Fig. 1(b), the signal is antisymmetric with respect to turning the sample and, thus, reverses completely in the 180°-configuration cap||F|N||sub. We note that we actually use this antisymmetric behavior to precisely turn the sample around. Very similar signals are observed for W as N material (Fig. 4).

**F sample.** The terahertz signals from the F sample [Fig. 2(b)] are two orders of magnitude smaller than from the F|N counterpart. When the sample is turned, the signal does not simply invert but changes shape. This behavior indicates a superposition of contributions which are symmetric (+) and asymmetric (−) under sample turning. To separate them, we calculate the signals

$$S_F^\pm(t) = \frac{S_F(t, \theta = 0°) \pm S_F(t, \theta = 180°)}{2}, \qquad (4)$$

which are displayed in Fig. 2(c). We emphasize that we can reliably reproduce $S_F^+(t)$ using two complementary approaches (see Appendix A and Note 1 in Supplemental Material).

The magnitude of the asymmetric component $S_F^-$ is comparable to that of $S_F^+$ and suggests that the F sample exhibits noticeable inversion asymmetry. This conclusion is not unexpected because thin films are known to exhibit inhomogeneities along the growth direction and to possess different properties at the substrate interface as compared to the bulk.[46]

The symmetric component $S_F^+(t)$ contains the contribution $E_M$ due to UDM [Fig. 1(a)]. Assuming that $S_F^+$ solely arises from $E_M$, we retrieve $E_M(t)$ and, thus, the evolution of the magnetization change $\Delta M(t)$ (see Section II). The extracted $\Delta M(t)$ is shown in Fig. 2(d) along with the magnetization dynamics measured by the transient MOKE [black curve in Fig. 2(d)]. The good agreement of the two curves in terms of sign, magnitude and shape is fully consistent with the notion that $S_F^+$ arises from UDM of the F sample.

**UDM vs TST.** We can now directly compare the terahertz signal waveforms $S_F^+(t)$ and $S_{F|N}^-(t)$ due to UDM of a single layer of F=CoFe [Fig. 1(a)] and TST from F into N [Fig. 1(b)]. The result is shown in Fig. 3(a) and reveals a remarkable correlation: The terahertz signals $S_F^+(t)$ and $S_{F|N}^-(t)$ exhibit completely identical dynamics. We emphasize that we make analogous observations for two other ferromagnets, F=CoFeB and NiFe [Fig. 3(a)], as well as N=W [Fig. 4(b)] and different film thicknesses.

Our observation $S_{F|N}^-(t) \propto S_F^+(t)$ and the origins of $S_{F|N}^-$ [Eq. (1)] and $S_F^+$ [Eq. (2)] imply that

$$j_s(t) \propto \dot{M}(t). \qquad (5)$$

In other words, our THz emission signals show directly that on ultrafast time scales, the photoinduced spin current in an F|N stack has a temporal evolution that is identical to that of the rate of photoinduced magnetization quenching of an F sample. The most explicit manifestation of Eq. (5) is Fig. 3(b), which shows the actual dynamics of $j_s$ and $\dot{M}$ as retrieved from the signals $S_{F|N}^-$ and $S_F^+$.

Equation (5) summarizes our central experimental result and reveals the profound relationship between UDM [Fig. 1(a)] and TST [Fig. 1(b)]. It strongly suggests that TST in an F|N stack and UDM of an F sample are driven by the same force.

**Driving force**

To identify this force, we consider the schematic of Fig. 1(c), which shows the density of states of spin-up (↑) and spin-down (↓) electrons vs single-electron energy $\epsilon$. We assume that UDM primarily arises from quasi-elastic spin flips[10] [white arrow in Fig. 1(c)] and that the pump pulse can be considered a small perturbation of the system. At a given $\epsilon$, the probability of a spin-flip event is proportional to the difference

$n^{F\uparrow}(\epsilon,t) - n^{F\downarrow}(\epsilon,t)$ where $n^{F\sigma}(\epsilon,t) = n_0(\epsilon) + \Delta n^{F\sigma}(\epsilon,t)$ denotes the occupation number of a Bloch state with spin $\sigma$ (↑ or ↓) and energy $\epsilon$ in F. It is a sum of the distribution $n_0$ of the unexcited sample and the pump-induced changes $\Delta n^{F\sigma}$. The rate $\dot{M}(t)$ of magnetization change is obtained by integrating over all energies $\epsilon$.

Similarly, the spin current $j_s(t)$ from F to N in the F|N stack is inferred by counting all spin transmission events across the F-N interface [Fig. 1(d)]. As detailed in the Appendix B, we find that

$$\left.\begin{array}{c}\dot{M}(t)\\j_s(t)\end{array}\right\} \propto \Delta\tilde{\mu}_s(t) + \text{(Seebeck terms)}, \tag{6}$$

where the quantity

$$\Delta\tilde{\mu}_s(t) = \int d\epsilon\, \left(n^{F\uparrow} - n^{F\downarrow}\right)(\epsilon,t) \tag{7}$$

has the same form for $\dot{M}(t)$ and $j_s(t)$, whereas the Seebeck contribution is different. Remarkably, Eq. (6) is fully consistent with our central experimental finding [Eq. (5)] if $\Delta\tilde{\mu}_s$ dominates. We, therefore, consider $\Delta\tilde{\mu}_s$ and the Seebeck terms in more detail.

If the occupation numbers $n^{F\sigma}$ in Eq. (7) are Fermi-Dirac functions with chemical potentials $\mu^{F\sigma}$, $\Delta\tilde{\mu}_s$ can be shown to equal the spin voltage[33,35] $\mu^{F\uparrow} - \mu^{F\downarrow}$. Therefore, $\Delta\tilde{\mu}_s$ can be considered a generalized spin voltage that is caused by an electron distribution with an arbitrary, possibly nonthermal imbalance $\Delta n^{F\uparrow} - \Delta n^{F\downarrow}$. Upon absorption of the pump pulse, $\Delta\tilde{\mu}_s$ rises immediately because spin-up and spin-down electrons in a Stoner-type ferromagnet possess a very different electronic density of states around the Fermi level [see Fig. 1(c) and Eq. (36)].

The Seebeck-type term[33] in Eq. (6) is proportional to the difference $\Delta\tilde{T}^{F\uparrow} - \Delta\tilde{T}^{F\downarrow}$ in the case of $\dot{M}(t)$, while it equals a linear combination of $\Delta\tilde{T}^{F\uparrow} - \Delta\tilde{T}^{N\uparrow}$ and $\Delta\tilde{T}^{F\downarrow} - \Delta\tilde{T}^{N\downarrow}$ for $j_s(t)$. Here, $\Delta\tilde{T}^{X\sigma}$ is the pump-induced change in the generalized temperature of electrons with spin $\sigma$ in X=F or N. It scales with the electronic excess energy [Eq. (40)] and equals the conventional temperature change once the electron distribution is thermal.

A comparison of our experimental results [Fig. 3 and Eq. (5)] with Eq. (6) strongly indicates that the Seebeck terms play a minor role in our photoexcited F and F|N samples, likely because all electronic subsystems X$\sigma$ attain approximately equal generalized temperatures faster than our time resolution of 40 fs. Consequently, we consider only one common generalized electron temperature $\Delta\tilde{T}^{X\sigma} = \Delta\tilde{T}_e$ in the following.

To conclude, our observations [summarized by Eq. (5)] and modeling [leading to Eq. (6)] directly imply that the generalized spin voltage $\Delta\tilde{\mu}_s$ of F is the dominant driving force of both UDM [Fig. 1(a)] and TST [Fig. 1(b)].

**Modeling the spin dynamics**

The identical temporal evolution of $\Delta\tilde{\mu}_s$ in the F and F|N samples shows that the coupling to N does not significantly perturb the dynamics of (i) spins and (ii) electrons in F. For a quantitative discussion, we relate the generalized spin voltage $\Delta\tilde{\mu}_s$ to the dynamics of the uniform generalized electron excess temperature $\Delta\tilde{T}_e$ (see Appendix B). We obtain

$$\Delta\tilde{\mu}_s(t) \propto \Delta\tilde{T}_e(t) - \Gamma_{es} \int_0^\infty d\tau\, e^{-\Gamma_{es}\tau}\, \Delta\tilde{T}_e(t-\tau), \tag{8}$$

where $\Gamma_{es}^{-1}$ is the time constant of electron-spin equilibration. To illustrate Eq. (8), we consider a step-like increase of the generalized uniform electron temperature. Once $\Delta\tilde{T}_e$ jumps to a nonzero value, $\Delta\tilde{\mu}_s(t)$ follows without delay according to the first term of Eq. (8). It triggers transfer of spin angular momentum from the F electrons into the F lattice (UDM) and, possibly, into N (TST). The loss of spin polarization, however, decreases $\Delta\tilde{\mu}_s$, which decays on the time scale $\Gamma_{es}^{-1}$, as dictated by the second term of Eq. (8).

In our experiment, the excess energy of the F electrons and, thus, $\Delta\tilde{T}_e$ rise instantaneously upon pump-pulse excitation, and they decay due to energy transfer to the crystal lattice.[47,48] As shown in Appendix C, we can accordingly model the evolution of $\Delta\tilde{T}_e$ by

$$\Delta\tilde{T}_e(t) \propto \Theta(t)[(1-R)e^{-\Gamma_{ep}t} + R], \tag{9}$$

where $\Theta(t)$ is the Heaviside step function, $\Gamma_{ep}^{-1}$ is the time constant of electron-phonon equilibration, and $R$ is the ratio of electronic and total heat capacity of the sample. With these assumptions, Eqs. (6) and (8) yield the simple result

$$\left.\begin{array}{c}\dot{M}(t)\\ j_s(t)\end{array}\right\} \propto \Theta(t)\left[A_{es}e^{-\Gamma_{es}t} - A_{ep}e^{-\Gamma_{ep}t}\right], \tag{10}$$

where $A_{es} = (\Gamma_{es} - R\Gamma_{ep})/(\Gamma_{es} - \Gamma_{ep})$ and $A_{ep} = (1-R)\Gamma_{ep}/(\Gamma_{es} - \Gamma_{ep})$. To account for our experimental time resolution, Eq. (10) is convoluted with a Gaussian of 40 fs full width at half maximum. We apply Eq. (10) to the measured $\dot{M}(t)$ and $j_s(t)$ (Fig. 4) and take only $\Gamma_{es}$ and the overall amplitude as free sample-dependent fit parameters. For $\Gamma_{ep}$ and $R$, literature values are assumed (see Appendix C and Table S2 in Supplemental Material). Figure 4 demonstrates that Eq. (10) excellently describes the measured $\dot{M}(t)$ and $j_s(t)$.

## IV. DISCUSSION

We can now discuss the impact of N on the dynamics of (i) the electron spins through $\Gamma_{es}$ and (ii) the electronic excess heat through $\Gamma_{ep}$. According to Eq. (10), the slope of $\dot{M}(t)$ and $j_s(t)$ approximately equals $-(\Gamma_{es} + \Gamma_{ep})$ right after excitation because both electron-spin and electron-phonon equilibration contribute to the decay dynamics. For an F sample with F=CoFeB, we find $\Gamma_{es}^{-1} = 104$ fs [Fig. 4(a)], which agrees with previous reports[49] and is four times smaller than $\Gamma_{ep}^{-1} = 420$ fs. Therefore, we have $\Gamma_{es} \gg \Gamma_{ep}$, and the slope of the initial decay of $\dot{M}(t)$ is dominated by $\Gamma_{es}$.

When N=Pt is attached to CoFeB, we expect a larger $\Gamma_{es}$ (due to TST) and an increase of $\Gamma_{ep}$ by 20% (see Table S2 in Supplemental Material). In contrast, we observe an equally fast decay of $\dot{M}(t)$ and $j_s(t)$ [Fig. 4(a)]. Therefore, the time constant $\Gamma_{es}^{-1}$ of the F|N sample does not decrease markedly as confirmed by our fits, which yield a very similar $\Gamma_{es}^{-1}$ for CoFeB and CoFeB|Pt. In other words, TST into the Pt layer does not accelerate spin-electron equilibration ($\Gamma_{es}$), and the slightly faster electron cooling ($\Gamma_{ep}$) is negligible because $\Gamma_{ep} \ll \Gamma_{es}$. This reasoning is fully confirmed by the almost identical spin-current dynamics $j_s(t)$ in CoFeB|W and CoFeB|Pt [Fig. 4(a)].

We finally test NiFe as F material because its $\Gamma_{es}$ is smaller than for CoFeB.[50] Indeed, both $\dot{M}(t)$ and $j_s(t)$ decay 50% more slowly for NiFe than for CoFeB [Fig. 4(b)] while $\Gamma_{es}$ remains the same for the NiFe|Pt and NiFe samples within our experimental uncertainty.

## V. CONCLUSIONS

Our experiments show that, following optical excitation, the rate of change $\dot{M}(t)$ of the magnetization of an F sample [Fig. 1(a)] and the spin current density $j_s(t)$ from F to an adjacent N [Fig. 1(b)] exhibit identical dynamics. According to our analysis, this behavior relies on two reasons.

First, UDM and TST are driven by the same force: a generalized spin voltage [Eq. (6)], which quantifies the urgency with which the magnetization of the ferromagnet aims to adapt to its instantaneous electronic temperature. We suggest to term the heat-induced spin voltage the pyrospintronic effect because it is analogous to the pyroelectric effect of a pyroelectric material in which the spontaneous electric polarization aims to follow the instantaneous temperature. Note, however, that our measured spin current is not the result of a spin-dependent Seebeck effect:[33] To quantitatively explain our data, we neither have to assume a temperature difference between F and N nor between majority and minority electrons in F.

Second, after the pump pulse has excited the electronic system of F, the generalized spin voltage and, thus, $\dot{M}(t)$ and $j_\mathrm{s}(t)$ jump to a nonzero value and subsequently relax by electron-spin equilibration, while the significantly slower electron-phonon equilibration has a minor influence. Our results also strongly suggest that the impact of TST on $\Gamma_\mathrm{es}$ is negligible in our experiments.

The last conclusion implies that also in the F|N stack, the photoinduced spin voltage primarily decays due to spin-flip processes in F. In other words, only a small fraction of the available spin angular momentum is transferred to N. We estimate that the spin-current amplitude can, in principle, be increased by one order of magnitude by using more transparent F-N interfaces and F materials with larger electron-spin relaxation time $\Gamma_\mathrm{es}^{-1}$.

Regarding speed and bandwidth, we note that the temporal onset of TST is truly ultrafast and only limited by the duration of the pump pulse depositing energy in the electrons of F [Eq. (8)]. This feature is in remarkable contrast to the interfacial spin Seebeck effect,[42] where carrier multiplication is required to reach maximum spin current.

Importantly, our study allows us to apply the extensive knowledge about UDM of F samples to TST from F to an adjacent layer N. This insight is expected to be very helpful to boost spin-current amplitudes in numerous applications such as spin torque,[22,23] spintronic terahertz emitters[24,25,26,27] and, potentially, energy harvesting.[51] Our findings also provide a new straightforward link between concepts of femtomagnetism and spintronics. In particular, terahertz emission spectroscopy holds great promise to be an excellent ultrafast monitor of the evolution of the generalized spin voltage.

## APPENDIX A: EXPERIMENTAL DETAILS

**Sample preparation and characterization**

The F samples (where F is $Co_{60}Fe_{20}B_{20}$, $Co_{70}Fe_{30}$ or $Ni_{80}Fe_{20}$) and F|N stacks (where N is Pt or W) are grown by means of magnetron sputtering. The deposition is performed at an Ar pressure of $4 \times 10^{-3}$ mbar at growth rates between 0.2 Å/s and 1 Å/s, depending on the material. Half of the substrate is covered by a metallic mask during deposition of the N material, thereby resulting in an F sample and an F|N stack on the same substrate and in the same run. All samples are protected by a 10 nm thick $Al_2O_3$ layer grown by atomic layer deposition. As substrates, we choose diamond and, for test purposes, fused silica.

Hysteresis loops show that the samples have a coercive field below 10 mT. We also measure the optical absorptance $A$ of the pump pulse and the sample impedance $Z$ from 1 to 7 THz as detailed in Ref. 52. We find that $Z$ is approximately independent of frequency. Values of $A$, $Z$ and the mean terahertz conductivity are compiled in Table S3 in Supplemental Material.

**Signals due to TST and UDM**

For the F|N stack, the signal is dominated by TST and the inverse spin Hall effect in N, which converts the electron spin current with density $(\hbar/2)j_s$ into a charge current with density $(-e)j_c$. Here, $\hbar$ is the Planck constant, $-e$ is the electron charge, and $j_c = \theta_{SH} j_s$ with $\theta_{SH}$ being the spin Hall angle of the N material. In the frequency domain, the terahertz electric field behind the sample is related to the spin current injected into the Pt layer by a generalized Ohm's law,[24]

$$E_{j_s}(\omega) = eZ(\omega)\theta_{SH}(\omega)\lambda_{\text{rel}} j_s(\omega). \qquad (11)$$

Terahertz transmission measurements and broadband measurements of the anomalous Hall effect of magnetic metals[40] show that the sample impedance $Z(\omega)$ and the spin Hall angle $\theta_{SH}(\omega)$ are constant over the relevant frequency range. Therefore, Eq. (11) yields $E_{j_s}(t) \propto j_s(t)$ in the time domain.

The time-dependent magnetization of the F sample gives rise to magnetic-dipole radiation with an electric field[20]

$$E_M(\omega) = -\frac{i\omega n d_F}{c} Z(\omega) M(\omega) \qquad (12)$$

directly behind the sample. Here, $n(\omega)$ is the refractive index of the half-space (substrate or cap window) that is not traversed by $E_M$, $d_F$ is the F thickness, and $c$ is the speed of light. Because the refractive index of our substrate and cap windows and the impedance $Z(\omega)$ are approximately independent of $\omega$ for the terahertz frequencies relevant here, Eq. (12) leads to $E_M(t) \propto \dot{M}(t)$ in the time domain.

**Measurement configurations**

*Symmetry considerations.* The terahertz emission signal from the F|Pt and F|W samples is dominated by the electric-dipole field $E_{j_s}$ [Fig. 1(b)]. In contrast, the terahertz magnetic-dipole field $E_M$ from the F sample [Fig. 1(a)] is typically two orders of magnitude smaller. It can easily be masked by spurious electric-dipole-type signals that arise when inversion symmetry is broken, either by the sample structure (structural inversion asymmetry, SIA) or by the perturbing light field (light-induced inversion asymmetry, LIA). For example, SIA can be caused by inequivalent interfaces of F,[45] and LIA can arise from a change of the pump intensity across the F thickness.[44]

To discriminate a terahertz electric-dipole field $E_{SIA}$ due to SIA from $E_M$, three different approaches are implemented. In the first approach, we symmetrize the sample by adding a cap layer (cap) that is identical to the substrate [sub; see inset of Fig. 2(a) and Fig. S1(a) in Supplemental Material]. We measure the sample both in the 0°-configuration sub||F||cap and the 180°-turned configuration cap||F||sub. While $E_{SIA}$ changes sign,[52] $E_M$ stays invariant. To minimize the field $E_{LIA}$ owing to LIA, which is also invariant under sample turning, we choose an F thickness much thinner than the attenuation length of the optical pump field (~30 nm). Calculations show that in our metal stacks, the pump field changes by less than 5% over the full thickness of up to 10 nm.

***Implementation.*** We acquire terahertz emission data from the F sample and the associated F|N stack, both of which are grown on the same substrate and can be reached by translating the sample perpendicularly to the pump beam. As sample spots, we choose F and F|N regions as close as possible to guarantee identical optical environments for the probed F and F|N thin-film regions. To reproducibly put the metal film into the focal region of the pump spot, we use crossed beams of alignment lasers to mark the position and tilt angle of the sample. To test for correct alignment, we check that the emission signals from the 0° and 180° sample configurations of the F|N sample are reversed versions of each other.

In the second approach, we measure unsymmetrized samples sub||F and sub||F|N analogous to the first method. Because of its macroscopic asymmetry, the signals from the 0° and 180° configurations of the sub||F|N sample are in general not reversed versions of each other anymore. The two signals, are, however, connected by a transfer function that can be easily inferred and, in turn, applied to the two signals from the sub||F sample. More details and another separation method working in reflection mode are presented in Note 1 of Supplemental Material. We emphasize that all three separation methods deliver highly consistent results.

### MOKE probing of magnetization dynamics

To interrogate the magnetization dynamics of the F sample by the magneto-optic Kerr effect (MOKE), we conduct a pump-probe measurement in which pump and probe pulses are incident onto the sample under 50° angle of incidence. Pump pulses (duration of 200 fs, center wavelength of 400 nm, repetition rate of 1 kHz) are obtained by frequency-doubling of pulses from a Ti:sapphire laser amplifier. Probe pulses (40 fs, 800 nm, 80 MHz) are taken from the seed oscillator of the amplifier.[53] During reflection off the sample, the probe polarization acquires an additional rotation and ellipticity, part of which is proportional to the sample magnetization averaged over the probing volume.

The pump-induced polarization variation of the reflected probe pulse is measured using a balanced detection scheme. In our samples, rotation and ellipticity signals have the same dynamics, indicating negligible pump-induced variation of magnetooptic constants. We confirm that the response is linear with respect to the used pump fluence of up to 1 mJ/cm². To push the time resolution down to 130 fs, the pump-probe transient is deconvoluted with the pump-pulse profile.

### APPENDIX B: SPIN-DYNAMICS MODEL

### Rate equations

Our goal is to model the spin dynamics of a single thin ferromagnetic metal layer F and an F|N stack where F is in contact with a thin normal-metal layer N. We assume that each layer X (F or N) can be treated as homogeneous and that the state of the electronic system in a given layer X is fully characterized by the occupation numbers $n_k^{X\sigma}$ of a Bloch state $(k, \sigma)$. Here, $\sigma = \uparrow, \downarrow$ refers to the electron spin, and $k$ summarizes the band index and wavevector. We define the magnetic moment $\boldsymbol{m} = m\boldsymbol{u}_y$ of F such that $(g^F/2)\mu_B m = MV^F$ where $\boldsymbol{M} = M\boldsymbol{u}_y$ is the magnetization [Fig. 1(a)], $g^F$ is the electron $g$-factor, $\mu_B$ is the Bohr magneton, and $V^F$ is the volume of F. Similarly, we define the spin current through the interface as $J_s = j_s A^F$ where $(\hbar/2)j_s$ is the spin-current density, and $A^{F|N}$ is the area of the F|N stack.

We adopt a simplified description in which the occupation of each Bloch state $(k, \sigma)$ is fully given by its energy $\epsilon_k^\sigma(t)$, that is,

$$n_k^{X\sigma}(t) = n^{X\sigma}(\epsilon_k^{X\sigma}(t), t). \tag{13}$$

To model magnetic order, we make use of the Stoner model, in which the Bloch energy depends on the pump-induced change $\Delta m$ in the magnetic moment according to

$$\epsilon_k^{X\sigma}(t) = \epsilon_{k0}^{X\sigma} + I^{X\sigma}\Delta m(t) + e\Phi^X(t). \tag{14}$$

Here, $\epsilon_{k0}^{X\sigma}$ is the Bloch energy before arrival of the pump pulse, and $I^{X\sigma} = I^{X\uparrow,\downarrow} = \pm I^X/2$ quantifies the strength of the effective electron-electron Coulomb interaction for X=F only. The electrostatic potential $\Phi^X$ accounts for a possible charging of a given layer X due to transport, where $-e$ is the electron charge.

Before arrival of the pump pulse, the $n^{X\sigma}(\epsilon,t)$ are given by one and the same Fermi-Dirac function $n_0(\epsilon)$ at temperature $T_0$. We now focus on the rate of change $\dot{n}^{F\sigma} = \partial n^{F\sigma}/\partial t$ of the electron occupation numbers $n^{F\sigma}$ in F. As detailed in the following, it is determined by four contributions,

$$\dot{n}^{F\sigma} = \dot{n}^{F\sigma}|_{sc} + \dot{n}^{F\sigma}|_{sf} + \dot{n}^{F\sigma}|_{tr} + \dot{n}^{F\sigma}|_{I}. \tag{15}$$

The first term on the right-hand side of Eq. (15) captures spin-conserving scattering events and the excitation by the pump pulse. It, thus, fulfills

$$0 = \int d\epsilon \, D^{F\sigma} \dot{n}^{F\sigma}|_{sc}, \tag{16}$$

where $D^{X\sigma}(\epsilon,t) = \sum_k \delta\left(\epsilon - \epsilon_k^{X\sigma}(t)\right)$ is the instantaneous density of Bloch states with spin $\sigma$.

Spin-flip events are captured by the second term of Eq. (15) and assumed to be quasi-elastic following Refs. 10 and 36. As indicated by Fig. 1(c), the rate of change of the electron occupation $n^{F\uparrow}$ due to elastic spin-flip scattering is proportional to $n^{F\uparrow}$ and the number $(1-n^{F\downarrow})D^{F\downarrow}$ of available unoccupied spin-down states at the same energy $\epsilon$ plus an analogous term for the reverse process,

$$\dot{n}^{F\uparrow}|_{sf} = -P_{sf}^F n^{F\uparrow}(1-n^{F\downarrow})D^{F\downarrow} + P_{sf}^F n^{F\downarrow} D^{F\downarrow}(1-n^{F\uparrow}) = -(n^{F\uparrow} - n^{F\downarrow})\frac{g_{sf}}{D^{F\uparrow}}. \tag{17}$$

Here, $g_{sf}(\epsilon) = (P_{sf}^F D^{F\uparrow} D^{F\downarrow})(\epsilon)$, and the factor $P_{sf}(\epsilon)$ is proportional to the square of the matrix element for a spin-flip scattering event. The analogous equation for the rate of change of $n^{\downarrow}(\epsilon)$ is obtained by simply swapping ↑ and ↓.

The third term of Eq. (15) captures spin transfer across the F-N interface [see Fig. 1(d)]. We assume the transmission events to be spin-conserving and elastic. Consequently, we can consider spin-up ($\sigma=\uparrow$) and spin-down ($\sigma=\downarrow$) electrons separately. By counting transmission events analogous to Eq. (17), we obtain

$$\dot{n}^{F\sigma}|_{tr} = -(n^{F\sigma} - n^{N\sigma})\frac{g_{tr}^\sigma}{D^{F\sigma}}, \tag{18}$$

where $g_{tr}^\sigma(\epsilon) = (T_{tr}^\sigma D^{F\sigma} D^{N\sigma})(\epsilon)$, and $T_{tr}^\sigma(\epsilon)$ is a spin-dependent interface transmittance.

The last term of Eq. (15) arises because $n$ is evaluated at a fixed $\epsilon$ while the Bloch energy changes according to Eq. (14). We obtain

$$\dot{n}^{F\sigma}|_I = n^{F\sigma'} I^{F\sigma} \dot{m} = I^{F\sigma}(\partial_\epsilon n^{F\sigma})(\dot{m}|_{sf} + \dot{m}|_{tr}), \tag{19}$$

where $n^{F\sigma'} = \partial n^{F\sigma}/\partial \epsilon$. In the last step of Eq. (19), we split the rate of change of the magnetization into the contributions of spin flips and spin transport. As the electronic band structure depends on the magnetic moment $m$ (see Eq. (14)), $D^{F\sigma}(\epsilon)$, $g_{sf}(\epsilon)$ and $g_{tr}^\sigma(\epsilon)$ are also time-dependent. This time dependence is left implicit in our discussion.

**Spin transfer rates**

We are interested in the dynamics of the F magnetic moment

$$m = \int d\epsilon \, (D^{F\uparrow} n^{F\uparrow} - D^{F\downarrow} n^{F\downarrow}). \tag{20}$$

Using Eq. (17), its rate of change due to spin-flip events is given by

$$\dot{m}|_{\text{sf}} = -2 \int d\epsilon \, (n^{\text{F}\uparrow} - n^{\text{F}\downarrow}) g_{\text{sf}}^{\text{F}}. \tag{21}$$

Using Eq. (18), the spin-resolved electron current through the F-N interface can be calculated by

$$J^\sigma = \int d\epsilon \, (n^{\text{F}\sigma} - n^{\text{N}\sigma}) g_{\text{tr}}^\sigma. \tag{22}$$

We note that Eqs. (21) and (22) yield zero spin transfer before the pump pulse arrives because in this case, all distribution functions $n^{\text{F}\sigma}$ and $n^{\text{N}\sigma}$ equal the same Fermi-Dirac distribution $n_0$ with chemical potential $\mu_0$ and temperature $T_0$.

**Moment expansion**

As the relevant observables $\dot{m}|_{\text{sf}}$ and $J^\sigma$ involve differences of distribution functions only, we focus our discussion on the difference

$$\Delta n^{\text{X}\sigma} = n^{\text{X}\sigma} - n_0 \tag{23}$$

of the distribution function $n^{\text{X}\sigma}(\epsilon, t)$ and the equilibrium distribution $n_0$. We assume that $\Delta n^{\text{X}\sigma}$ is significantly nonzero only in a relatively narrow energy window around the chemical potential $\mu_0$ of the unperturbed system and that the energy-dependent weight factors $D^{\text{F}\sigma}(\epsilon)$, $g_{\text{sf}}(\epsilon)$ and $g_{\text{tr}}^\sigma(\epsilon)$ can be well approximated by the Sommerfeld approximation

$$W(\epsilon) \approx W(\mu_0) + W'(\mu_0)(\epsilon - \mu_0), \tag{24}$$

where $W$ stands for $D^{\text{F}\sigma}$, $g_{\text{sf}}$ or $g_{\text{tr}}^\sigma$. Integrals involving these functions, such as Eqs. (21) and (22), then turn into

$$\int d\epsilon \, W(\epsilon) \Delta n^{\text{X}\sigma}(\epsilon) = W(\mu_0) \Delta P^{\text{X}\sigma} + W'(\mu_0) \Delta A^{\text{X}\sigma}, \tag{25}$$

which is just a linear combination of the zeroth and first moment of $\Delta n^\sigma$, that is,

$$\Delta P^{\text{X}\sigma} = \int d\epsilon \, \Delta n^{\text{X}\sigma} \quad \text{and} \quad \Delta A^{\text{X}\sigma} = \int d\epsilon \, (\epsilon - \mu_0) \Delta n^{\text{X}\sigma}. \tag{26}$$

In the case that $n^{\text{X}\sigma} = n_0 + \Delta n^{\text{X}\sigma}$ is a Fermi-Dirac distribution with chemical potential $\mu^{\text{X}\sigma}$ and temperature $T^{\text{X}\sigma}$, the $\Delta P^{\text{X}\sigma}$ and $\Delta A^{\text{X}\sigma}$ become Fermi-Dirac integrals and reduce to

$$\Delta P^{\text{X}\sigma} = \mu^{\text{X}\sigma} - \mu_0 \quad \text{and} \quad \Delta A^{\text{X}\sigma} = \frac{\pi^2 k_B^2}{6}[(T^{\text{X}\sigma})^2 - T_0^2] + \frac{1}{2}(\mu^{\text{X}\sigma} - \mu_0)^2. \tag{27}$$

Because $(\mu^{\text{X}\sigma} - \mu_0)^2$ is typically small, one can interpret $\Delta P^\sigma$ and $\Delta A^\sigma$, respectively, as changes in a generalized chemical potential and a squared generalized temperature. We emphasize, however, that the definition of the moments $\Delta P^{\text{X}\sigma}$ and $\Delta A^{\text{X}\sigma}$ [Eq. (26)] also applies to nonthermal electron distributions $n_0 + \Delta n^{\text{X}\sigma}$.

In Ref. 35, the difference $\Delta \mu_s = \Delta \mu^{\text{F}\uparrow} - \Delta \mu^{\text{F}\downarrow}$ is termed spin voltage. We accordingly term

$$\Delta P_s = \Delta P^{\text{F}\uparrow} - \Delta P^{\text{F}\downarrow} \tag{28}$$

generalized spin voltage. In the main text, $\Delta P_s$ is written as $\Delta \tilde{\mu}_s$, and further below [Eq. (38)], we will express $\Delta A^{\text{X}\sigma}$ by the generalized excess temperature $\Delta \tilde{T}^{\text{X}\sigma}$ of the X$\sigma$ electrons.

As the pump-induced variation of the electron distribution functions $n^{\text{X}\sigma}$ and, thus, the transient state of the electronic system are fully characterized by the moments $\Delta P^{\text{X}\sigma}$ and $\Delta A^{\text{X}\sigma}$, it is sufficient to determine the dynamics of $\Delta P^{\text{X}\sigma}$ and $\Delta A^{\text{X}\sigma}$. This conclusion is consistent with a recent thermodynamic treatment of ultrafast spin dynamics.[37] In the following, we will connect the phenomenological coupling coefficients of Ref. 37 with the parameters of our simplified microscopic description.

**Relevant observables**

We apply Eq. (25) to the rate of change of the magnetic moment [Eq. (21)]. We find

$$\dot{m}|_{sf} = -2g_{sf}(\mu_0)\Delta P_s - 2g'_{sf}(\mu_0)(\Delta A^{F\uparrow} - \Delta A^{F\downarrow}), \qquad (29)$$

where the first term on the right-hand side describes magnetization relaxation driven by the generalized spin voltage [Eq. (28)]. The term proportional to $\Delta A^{F\uparrow} - \Delta A^{F\downarrow}$ is a term analogous to the Seebeck effect, which contributes as long as the generalized temperatures of spin-up and spin-down electrons are different.

The magnetic moment of F is also modified by spin transport through the F-N interface, $-\dot{m}|_{tr} = J_s = J^{\uparrow} - J^{\downarrow}$. We assume vanishing charge transport, $J^{\uparrow} + J^{\downarrow} = 0$, and the same chemical potential for spin-up and spin-down electrons in N, $\Delta P^{N\uparrow} = \Delta P^{N\downarrow} = \Delta P^N$. These assumptions allow us to eliminate $\Phi^N - \Phi^F$ (see Note 2 of Supplemental Material). Along with Eqs. (22), (25) and (26), we find

$$-\dot{m}|_{tr} = J_s = g_{tr}(\mu_0)\Delta P_s + s^{\uparrow}_{tr}(\mu_0)(\Delta A^{F\uparrow} - \Delta A^{N\uparrow}) - s^{\downarrow}_{tr}(\mu_0)(\Delta A^{F\downarrow} - \Delta A^{N\downarrow}), \qquad (30)$$

where $2g_{tr}^{-1} = (g_{tr}^{\uparrow})^{-1} + (g_{tr}^{\downarrow})^{-1}$ and $s^{\sigma}_{tr} = g_{tr}g^{\sigma\prime}_{tr}/g^{\sigma}_{tr}$. The two final terms in Eq. (30) are again of Seebeck type and vanish once the temperatures of F and N have equilibrated. In this regime, the driving force of both $\dot{m}^F|_{sf}$ and $\dot{m}^F|_{tr}$ is given solely by the spin voltage $\Delta P_s$ of F.

The total energy of the F electrons is in the Stoner model given by $E^F = \sum_\sigma \int d\epsilon\, (\epsilon - \mu_0) D^{F\sigma} n^{F\sigma} + I^F m^2/4$. By using $\dot{D}^{F\sigma}(\epsilon) = D^{F\sigma\prime}(\epsilon) I^{F\sigma} \dot{m}$ and Eq. (20), we find that the rate of change obeys

$$\dot{E}^F = \sum_\sigma \int d\epsilon\, (\epsilon - \mu_0) D^{F\sigma} (\dot{n}^{F\sigma} - \dot{n}^{F\sigma}|_I), \qquad (31)$$

where the term $\dot{n}^{F\sigma}|_I$ [Eq. (19)] takes the time-dependence of the Bloch energies into account.

**Time evolution of $\Delta P_s$**

To determine the dynamics of the system and, thus, the magnetization, it is sufficient to determine the dynamics of the moments, that is, the generalized spin voltage $\Delta P_s$ and squared temperatures $\Delta A^{X\sigma}$. According to Eqs. (15) and (19), we need to consider contributions of spin flips, spin transport and spin-conserving processes,

$$\Delta \dot{P}_s = \Delta \dot{P}_s|_{sf} + \Delta \dot{P}_s|_{tr} + \Delta \dot{P}_s|_{sc}. \qquad (32)$$

By taking the time derivative of Eq. (26), considering Eqs. (17) and (19), performing the moment expansion of Eq. (25), and using Eq. (29), we obtain (see Note 2 of Supplemental Material)

$$\Delta \dot{P}_s|_{sf} = -\frac{2}{\chi^F(\mu_0)}\left[g_{sf}(\mu_0)\Delta P_s + s_{sf}(\mu_0)(\Delta A^{F\uparrow} - \Delta A^{F\downarrow})\right], \qquad (33)$$

where $1/\chi^F = (1/D^{F\uparrow} + 1/D^{F\downarrow})/2 - I^F$ is the inverse of the Pauli susceptibility $\chi^F = \partial m/\partial \mu_s$ of F, and $s_{sf} = g'_{sf} - \chi^F g_{sf}\left[D^{F\uparrow\prime}/(D^{F\uparrow})^2 + D^{F\downarrow\prime}/(D^{F\downarrow})^2\right]/2$ is the coefficient describing the Seebeck-type response of the spin voltage to a temperature difference between majority and minority electrons.

To determine the contribution of spin transport, we take the time derivative of Eq. (26), consider Eqs. (18) and (19), perform the moment expansion of Eq. (25) and use Eq. (30). Making the same assumptions as in the derivation of Eq. (30), we obtain (see Note 2 of Supplemental Material)

$$\Delta \dot{P}_s|_{tr} = -\frac{1}{\chi^F(\mu_0)}\left[g_{tr}(\mu_0)\Delta P_s + \tilde{s}^{\uparrow}_{tr}(\mu_0)(\Delta A^{F\uparrow} - \Delta A^{N\uparrow}) - \tilde{s}^{\downarrow}_{tr}(\mu_0)(\Delta A^{F\downarrow} - \Delta A^{N\downarrow})\right], \qquad (34)$$

where $\tilde{s}^{\sigma}_{tr} = s^{\sigma}_{tr} - g^{\sigma\prime}_{tr}\chi^F/D^{F\sigma}$.

Excitation by the pump pulse and subsequent spin-conserving electron-electron and electron-lattice interactions also affect the occupation numbers $n^{X\sigma}$. By applying the moment expansion of Eq. (25) to Eq. (16), we find that spin-conserving scattering processes couple the spin voltage and the generalized temperature through

$$\Delta \dot{P}_s \big|_{sc} = -\frac{D^{F\uparrow'}}{D^{F\uparrow}}(\mu_0) \Delta \dot{A}^{F\uparrow} \big|_{sc} + \frac{D^{F\downarrow'}}{D^{F\downarrow}}(\mu_0) \Delta \dot{A}^{F\downarrow} \big|_{sc}. \tag{35}$$

Equation (32) along with Eqs. (33), (34) and (35) determine the dynamics of the spin voltage, provided the dynamics of the squared generalized temperatures $\Delta A^{X\sigma}$ are given. In these equations, the prefactors of $\Delta A^{X\sigma}$ and $\Delta P_s$ depend on the instantaneous state of the system and, thus, on the time-dependent occupation numbers $n^{X\sigma} = n_0 + \Delta n^{X\sigma}$.

**Linear excitation limit**

From now on, we focus on the limit of weak optical excitation of the F and F|N samples. In fact, in our experiments, all terahertz emission signals were found to scale linearly with the incident pump-pulse energy up to the maximum available incident fluence of 0.2 mJ/cm². Therefore, $\dot{m}$ and $J_s$ and, through Eqs. (29) and (30), $\Delta P_s$ and $\Delta A^{X\sigma}$, and, by Eq. (26), the changes in the occupation numbers $\Delta n^{X\sigma}$ are also directly proportional to the deposited pump power. It follows that the prefactors in Eqs. (33), (34) and (35) are independent of the pump-induced changes $\Delta n^{X\sigma}$ in the occupation numbers and can, thus, be evaluated for the unperturbed system. This simplification has important consequences.

First, we can solve Eq. (32) along with Eqs. (33), (34) and (35) for the spin voltage $\Delta P_s$. We find that $\Delta P_s$ is a convolution

$$\Delta P_s(t) = -(H_{es} * \Delta F)(t) = -\int d\tau\, H_{es}(t-\tau) \Delta F(\tau) \tag{36}$$

of a driving force $\Delta F$ with a response function $H_{es}(t) = \Theta(t) e^{-\Gamma_{es} t}$, where $\Theta(t)$ is the Heaviside step function. The exponential decay rate equals $\Gamma_{es} = 2g_{sf}/\chi^F$ for the F sample and $\Gamma_{es} = (2g_{sf} + g_{tr})/\chi^F$ for the F|N stack. The expression for the driving force $\Delta F$ is

$$\Delta F = \frac{D^{F\uparrow'}}{D^{F\uparrow}} \Delta \dot{A}^{F\uparrow}\big|_{sc} - \frac{D^{F\downarrow'}}{D^{F\downarrow}} \Delta \dot{A}^{F\downarrow}\big|_{sc} + \frac{s_{sf}}{\chi^F}(\Delta A^{F\uparrow} - \Delta A^{F\downarrow}) + \frac{\tilde{s}_{tr}^{\uparrow}}{\chi^F}(\Delta A^{F\uparrow} - \Delta A^{N\uparrow}) - \frac{\tilde{s}_{tr}^{\downarrow}}{\chi^F}(\Delta A^{F\downarrow} - \Delta A^{N\downarrow}), \tag{37}$$

where all prefactors should be evaluated at $\epsilon = \mu_0$ and for the unperturbed system. The first two terms of $\Delta F$ cause a change in the spin voltage, and they scale with the time derivative of the pump-induced excess energy of spin-up and spin-down electrons. The remaining terms are Seebeck-type terms that disappear when the generalized temperatures of all electron subsystems $X\sigma$ have the same value. The last two terms in Eq. (37) are omitted for the case of an F sample.

Second, the pump-induced change in the squared generalized temperature [Eq. (27)] of electron system $X\sigma$ simplifies to

$$\Delta A^{X\sigma} = \frac{\pi^2 k_B^2}{3} T_0 \Delta \tilde{T}^{X\sigma}, \tag{38}$$

where $T_0 + \Delta \tilde{T}^{X\sigma}$ can be interpreted as generalized temperature of the $X\sigma$ electrons. The expression for the generalized chemical potential $\mu_0 + \Delta P^{X\sigma} = \mu_0 + \Delta \tilde{\mu}^{X\sigma}$ remains unchanged.

Third, the rate of change of the energy of the X-layer electrons (see Eq. (31)) simplifies to

$$\dot{E}^F = \sum_\sigma \int d\epsilon\, (\epsilon - \mu_0) D_0^{F\sigma} \dot{n}^{F\sigma} = \sum_\sigma C_e^{F\sigma} \partial_t \Delta \tilde{T}^{F\sigma} \tag{39}$$

where $C_e^{X\sigma} = (\pi^2 k_B^2/3) T_0 D_0^{X\sigma}(\mu_0)$, and $C_e^X = C_e^{X\uparrow} + C_e^{X\downarrow}$ is the heat capacity of the X electrons. Here, we neglected terms of order $(\epsilon - \mu_0)^2$ in the spirit of the moment expansion of Eq. (25). Therefore, the excess energy of the F electrons is

$$\Delta E^F = \sum_\sigma C_e^{X\sigma} \Delta \tilde{T}^{X\sigma}, \quad (40)$$

which underscores the interpretation of $T_0 + \Delta \tilde{T}^{X\sigma}$ as generalized temperature.

**Dynamics of excess energy**

Owing to Eqs. (29), (30), (36) and (37), the dynamics of UDM and TST are fully determined by a linear combination of the $\Delta A^{X\sigma}$ and, because of Eq. (38), the generalized excess temperatures $\Delta \tilde{T}^{X\sigma}$ of all electron subsystems $X\sigma$.

To develop a simple model for the time dependence of the generalized temperature, we briefly review the processes following photoexcitation of metal thin films.[47] At time $t = 0$, the $\delta$-like pump pulse excites the sample, thereby causing a step-like increase of the electronic excess energy and, thus, of all $\Delta \tilde{T}^{X\sigma}$.

Due to electron-electron interactions, all electronic subsystems $X\sigma$ quickly reach thermal equilibrium with each other, resulting in approximately equal generalized electronic temperatures, $\Delta \tilde{T}^{X\sigma} = \Delta \tilde{T}_e$. In this limit, the Seebeck-type contributions to the magnetization dynamics [Eqs. (29) and (30)] and to the driving force $\Delta F$ [Eq. (37)] are absent. Because we do not observe any signature Seebeck-type terms in our experiment, we assume one uniform electron temperature ($\Delta \tilde{T}^{X\sigma} = \Delta \tilde{T}_e$) at all times. As a consequence, Eqs. (36), (37) and (38) result in Eq. (8) of the main text. As mentioned above, carrier multiplication is not relevant for modifying the excess energy and, thus, $\Delta \tilde{T}^{X\sigma}$.

Electron-phonon interaction, on the other hand, causes heat transfer from the electrons to the crystal lattice with time constant $\Gamma_{ep}^{-1}$. On a much longer time scale, which is not considered here, heat is transferred into the sample environment. Consequently, we model the time dependence of the generalized temperature by the ansatz

$$\Delta \tilde{T}_e(t) = \Theta(t)\left[\Delta \tilde{T}_\infty + (\Delta \tilde{T}_{e0} - \Delta \tilde{T}_\infty) e^{-\Gamma_{ep} t}\right]. \quad (41)$$

Here, $\Delta \tilde{T}_{e0}$ is the increase of the uniform generalized temperature after absorption of the $\delta(t)$-like pump pulse and the fast equilibration between all electron subsystems $X\sigma$. The term $\Delta \tilde{T}_\infty = R \Delta \tilde{T}_{e0}$ is the generalized excess temperature at which the combined electron and lattice system equilibrate, with $R$ being the ratio of electronic and total heat capacity. The driving force $\Delta F$ for the spin voltage then assumes the simple form

$$\Delta F = \gamma \partial_t \Delta \tilde{T}_e = \gamma \Delta \tilde{T}_{e0} \left[\delta(t) - (1-R)\Gamma_{ep} \Theta(t) e^{-\Gamma_{ep} t}\right], \quad (42)$$

where we abbreviated $\gamma = (\pi^2 k_B^2 T_0/3)\left(D^{F\uparrow\prime}/D^{F\uparrow} - D^{F\downarrow\prime}/D^{F\downarrow}\right)_0(\mu_0)$. Using Eq. (36), one immediately finds that

$$\Delta P_s(t) = -\gamma \Delta \tilde{T}_{e0} \Theta(t) \left[\frac{\Gamma_{es} - R\Gamma_{ep}}{\Gamma_{es} - \Gamma_{ep}} e^{-\Gamma_{es} t} - \frac{(1-R)\Gamma_{ep}}{\Gamma_{es} - \Gamma_{ep}} e^{-\Gamma_{ep} t}\right]. \quad (43)$$

Without the Seebeck-type contributions, $\dot{m}|_{sf}$ [Eq. (29)] and $\dot{m}|_{tr} = -j_s$ [Eq. (30)] are both directly proportional to $\Delta P_s(t)$, and Eq. (43) turns into Eq. (10) of the main text. To account for the time resolution of our experiment, we convolute Eq. (43) with a Gaussian of 40 fs full width at half maximum.

To fit our data with Eq. (43), we obtain $\Gamma_{ep}$ and $R$ from previous works and Eqs. (50) and (51). Prior to fitting, all measured curves are shifted to the same time zero. The only fit parameters are $\Gamma_{es}$ and an overall scaling factor. As seen in Fig. 4, we obtain excellent agreement with our measurements. All parameters and references are summarized in Table S1 of Supplemental Material.

## APPENDIX C: TWO-TEMPERATURE MODEL FOR NONTHERMAL STATES

To determine $\Gamma_{ep}$ and $R$ for an F sample, we extend the standard two-temperature model[47,48] (2TM) to nonthermal electron and phonon distributions and, subsequently, to a two-layer stack F|N.

### 2TM for F

To model the decay of the electronic excess heat in the F sample, we assume that equilibration between electron baths of different spins is much faster than electron-phonon equilibration. Therefore, all electron baths X$\sigma$ can be described by one common generalized excess temperature $\Delta\tilde{T}_e = \Delta\tilde{T}^{X\sigma}$, consistent with the notion of negligible Seebeck terms in Eq. (37).

Changes in the total electron energy of F arise from excitation by the pump laser and by energy transfer to the phonons. Using Eq. (39), the rate of change of the electron excess energy can, thus, be written as

$$\Delta\dot{E}^F = C_e^F \partial_t \Delta\tilde{T}_e = \Delta\dot{E}^F\big|_{ep} + \Delta\dot{E}^F\big|_{pump} \tag{44}$$

where $C_e^F = C_e^{F\uparrow} + C_e^{F\downarrow}$ is the total electronic heat capacity of F. The pump action is modeled as $\Delta\dot{E}^F\big|_{pump} = C_e^F \Delta\tilde{T}_{e0} \delta(t)$. To describe electron-phonon relaxation, we neglect spin flips and use[48]

$$\Delta\dot{E}^F\big|_{ep} \propto \sum_\sigma \int d\delta\, (\alpha^2 F)^{F\sigma}(\delta) \int d\epsilon\, \{[n^{F\sigma}(\epsilon) - n^{F\sigma}(\epsilon+\delta)]p^F(\delta) - [1 - n^{F\sigma}(\epsilon)]n^{F\sigma}(\epsilon+\delta)\}. \tag{45}$$

Here, $(\alpha^2 F)^{F\sigma}(\delta)$ denotes the Eliashberg function that describes the coupling of phonons of energy $\delta$ with two electronic states of the same spin $\sigma$ and energy $\epsilon$ and $\epsilon + \delta$. The occupation number of the phonons is given by $p(\delta)$. Note that the term under the $\epsilon$-integral becomes zero for all $\delta$ and $\epsilon$ provided $n^{F\sigma}$ is a Fermi-Dirac distribution and $p$ is a Bose-Einstein distribution with the same temperature.

By linearizing Eq. (45) with respect to $\Delta n^{F\sigma} = n^{F\sigma} - n_0$ and $\Delta p^F = p^F - p_0$, we obtain

$$\Delta\dot{E}^F\big|_{ep} \propto \sum_\sigma \int d\delta\, (\alpha^2 F)^{F\sigma}(\delta) \delta \Delta p^F(\delta)$$
$$- \sum_\sigma \int d\epsilon\, \Delta n^{F\sigma}(\epsilon) \int d\delta\, (\alpha^2 F)^{F\sigma}[1 - n_0(\epsilon-\delta) - n_0(\epsilon+\delta)]. \tag{46}$$

Because the weight factor of $\Delta n^{F\sigma}(\epsilon)$ in Eq. (46) is sufficiently smooth, it is legitimate to apply the moment expansion of Eq. (25), resulting in

$$\Delta\dot{E}^F\big|_{ep} \propto \sum_\sigma \int d\delta\, (\alpha^2 F)^{F\sigma}(\delta) \delta \Delta p^F(\delta) - \Delta A^F \sum_\sigma \int d\delta\, (\alpha^2 F)^{F\sigma}(\delta)[-2n_0'(\mu_0 - \delta)]. \tag{47}$$

The first integral approximately scales with the pump-induced phonon excess energy because $(\alpha^2 F)^{F\sigma}(\delta)$ is approximately proportional to the phonon density of states.[48] Owing to Eq. (40), the second integral approximately scales with the excess energy of the F electrons. The generalized chemical potential does not show up in Eq. (47) as the weight factor of $\Delta n^\sigma(\epsilon)$ in Eq. (46) is antisymmetric with respect to $\epsilon - \mu_0$.

When we finally assume that the phonon distribution $p_0 + \Delta p$ is thermal and obeys Bose-Einstein statistics at temperature $T_0 + \Delta T_p^F$, Eq. (47) leads to the familiar result

$$\Delta\dot{E}^F\big|_{ep} = -G_{ep}^F \cdot \left(\Delta\tilde{T}_e^F - \Delta T_p^F\right). \tag{48}$$

Here, the coupling strength $G_{ep}^F$ is proportional to $\sum_\sigma \int d\delta \, (\alpha^2 F)^{F\sigma}(\delta)[-2n_0'(\mu_0 - \delta)]$. In the last step to Eq. (48), we took advantage of the fact that $\Delta \dot{E}^F|_{ep} = 0$ when $\Delta \tilde{T}_e^F = \Delta T_p^F$. Equation (48) is the generalization of the 2TM to nonthermal electron distributions in the linear excitation limit.

To close the system of equations, an equation of motion for the phonon temperature analogous to Eqs. (44) and (48) is given by

$$C_p^F \partial_t \Delta T_p^F = +G_{ep}^F \cdot (\Delta \tilde{T}_e^F - \Delta T_p^F), \tag{49}$$

where $C_p^F$ is the phonon heat capacity of F.

**2TM for F|N stack**

To model the decay of the electronic excess heat in the F|N stack, we assume that equilibration between electron baths of different spins and in different layers is much faster than electron-phonon equilibration. Therefore, all electron baths X$\sigma$ can be described by one common generalized excess temperature $\Delta \tilde{T}_e = \Delta \tilde{T}^{X\sigma}$. The phonon bath of each layer couples to the electrons of the same layer. Direct coupling of phonons between F and N is neglected. The energy-flow diagram, the differential equations (analogous to Eqs. (44), (48) and (49)) and their solution are detailed in Note 2 of Supplemental Material.

We find that for the time scales relevant to our experiment, the dynamics of the generalized electron excess temperature is given by Eq. (41) with

$$\Gamma_{ep} = \frac{G_{ep}^F + G_{ep}^N}{C_e^F + C_e^N} \tag{50}$$

and

$$R = \frac{C_e^F + C_e^N}{C_e^F + C_e^N + C_p^F + C_p^N}. \tag{51}$$

Here, $C_e^X$ and $C_p^X$ are the heat capacities of electrons and phonons in X, respectively, and $G_{ep}^X$ quantifies electron-phonon coupling in X. For an F sample, the parameters $\Gamma_{ep}$ and $R$ are obtained by setting $C_e^N = 0$ and $G_{ep}^N = 0$ in Eqs. (50) and (51).

Note that the $C_e^X$, $C_p^X$ and $G_{ep}^X$ are extensive quantities because they refer to the F and N volumes that are effectively coupled to each other in terms of ultrafast energy exchange. For our stack geometry, we assume equal coupling lengths into the depth of F and N. Therefore, we can replace the extrinsic quantities $C_e^X$, $C_p^X$ and $G_{ep}^X$ by their specific (volume-normalized) counterparts, which can be obtained from literature (see Table S1 of Supplemental Material).

**Acknowledgments**

We thank Dr. Yuta Sasaki for contributions in an early stage of this work. We acknowledge funding by the German Science Foundation through the collaborative research center SFB TRR 227 "Ultrafast spin dynamics" (projects A05, B01, B02 and B03) and the priority program SPP 1666 "Topological Insulators" and funding by the European Union through the ERC H2020 CoG project TERAMAG/Grant No. 681917. We also acknowledge support by the International Max Planck Research School (IMPRS) for Elementary Processes in Physical Chemistry.